# Formal Verification of a Generic Algorithm for TDM Communication Over Inter Satellite Links

Miroslav Popovic
*University of Novi Sad*
*Faculty of Technical Sciences*
Novi Sad, Serbia
miroslav.popovic@rt-rk.uns.ac.rs

Marko Popovic
*RT-RK Institute for Computer Based Systems*
Novi Sad, Serbia
marko.popovic@rt-rk.com

Pavle Vasiljevic
*University of Novi Sad*
*Faculty of Technical Sciences*
Novi Sad, Serbia
pavle.vasiljevic@uns.ac.rs

Miodrag Djukic
*University of Novi Sad*
*Faculty of Technical Sciences*
Novi Sad, Serbia
miodrag.djukic@uns.ac.rs

***Abstract—The Python Testbed for Federated Learning Algorithms is a simple FL framework targeting edge systems, which provides the three generic algorithms: the centralized federated learning, the decentralized federated learning, and the universal TDM communication in the current time slot. The first two were formally verified in a previous paper using the CSP process algebra, and in this paper, we use the same approach to formally verify the third one, in two phases. In the first phase, we construct the CSP model as a faithful representation of the real Python code. In the second phase, the model checker PAT automatically proves correctness of the third generic algorithm by proving its deadlock freeness (safety property) and successful termination (liveness property).***

## I. INTRODUCTION

Python Testbed for Federated Learning Algorithms (PTB-FLA) [1], [2] is a simple Federated Learning (FL) framework that is being developed within the ongoing EU Horizon 2020 project "Trustworthy and Resilient Decentralised Intelligence for Edge Systems" (TaRDIS) [3]. PTB-FLA is generally targeting edge systems, and a particular edge system considered in this paper is a satellite constellation where satellites use federated learning to aid their navigation (i.e., orbit determination and propagation).

The PTB-FLA API is based on the Single Program Multiple Data (SPMD) pattern, and it provides the three generic algorithms: (i) the centralized federated learning, (ii) the decentralized federated learning, and (iii) the universal Time Division Multiplexing (TDM) communication (i.e., peer data exchange), in the current time slot, that may be used e.g., for orbit determination and time synchronization (ODTS) in LEO satellite constellations [4], [5], [6].

Since PTB-FLA is targeting safety and mission critical applications, all its generic algorithms must be formally verified thus granting its trustworthiness and enabling development of correct-by-construction (or correct-by-composition) PTB-FLA applications. The first two generic algorithms (centralized FL and decentralized FL) were formally verified in a 2-phases process by using the process algebra Communicating Sequential Processes (CSP) and the model checker Process Analysis Toolkit (PAT) [5].

In this paper, we use the same approach to formally verify the correctness of the third generic algorithm for the universal TDM communication, using the CSP process algebra and model checker PAT, in 2-phases process.

In the first phase, we construct by hand the CSP model of the generic algorithm as the faithful representation of the real Python code. We construct these models in a bottom-up fashion in two steps. In the first step, we construct the process corresponding to a generic algorithm instance, and in the second step, we construct the system model as an asynchronous interleaving of algorithm instances.

In the second phase, we formally verify CSP model constructed in the previous phase in two steps. In the first step, we formulate desired system properties, namely deadlock freeness (safety property) and successful FLA termination (liveness property). We formulate the latter property in two forms (the reachability statement and the always-eventually LTL formula). In the second step, we use PAT to automatically prove formulated verification statements.

The main contributions of this paper are: (i) the CSP model of the generic algorithm for universal TDM communication, and (ii) the formulations of the requested correctness properties. To the best of our knowledge, this is the first paper that formally verifies a generic algorithm of this type. However, this originality comes at the expense of a closely related work of other authors, which despite our best effort we could not find.

The paper is organized as follows. Section II briefly introduces the generic algorithm for universal TDM communication. Section III presents the CSP formal model. Section IV presents formal verification in PAT. Section V presents the more broadly related work. Section VI concludes the paper.

## II. THE UNIVERSAL TDM COMMUNICATION

The generic algorithm for the universal TDM communication was designed to support inter satellite communication in a satellite constellation, where an individual satellite has (i) an arbitrary number of antennas (note that different satellites may have different numbers of antennas) and (ii) an arbitrary number of peers (note that the number of peers is less or equal to the number of antennas).

Interestingly, for the current time slot in a TDM multiplex, a mathematical model of this generic algorithm is an algebraic relation, which is symmetric and anti-reflexive, and therefore may be represented by a graph whose vertices are PTB-FLA application instances (residing in satellites) and edges are the communication links among the instances. For example, in the GMV use case, under development in TaRDIS [3], this graph is typically a bus (see Fig. 1a) or a ring (see Fig. 1b).

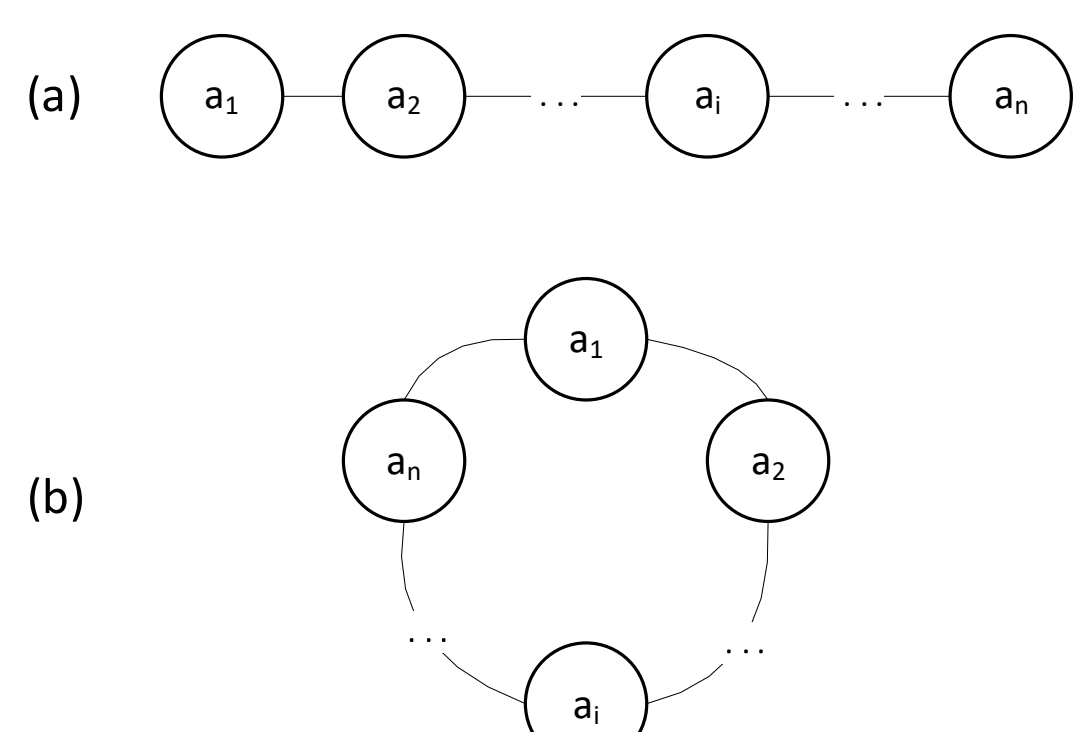

Fig. 1. The examples of typical TDM communication graphs.

The generic algorithm for the universal TDM communication was implemented as the function getMeas of the class PtbFla, see the module ptbfla_pkg/ptbfla.py at [2]. The next section presents the CSP formal model, where a node of a TDM communication graph is modeled as a CSP process and the complete TDM communication graph is modeled as an asynchronous interleaving of the processes.

## III. The CSP Formal Model

The next sections present: (i) the constants, variables, and channels, (ii) the process GetMeas, and (iii) the system.

### *A. The Constants, Variables, and Channels*

To parametrize the model, the following constants are introduced: NoNodes is the number of application instances, NoNeighbors is the maximum number of instances to communicate with, and NoTimeSlots is the number of time slots in the TDM multiplex.

To provide storage for input/output data and system state, the following variables are introduced: *terminated* is an indicator initially equal to False and set to True once the application terminates, *idataArr*[NoNodes] is a 1D array of input data whose elements correspond to individual nodes, *odataArr*[NoNodes][NoNodes-1] is a 2D array of output data whose rows are data received by individual nodes, *timeSlot*[NoNodes] is a 1D of the current time slot numbers whose elements correspond to individual nodes, *peerIds*[NoNodes][NoNodes][NoTimeSlots] is a 3D array whose element with indexes ($i$, $j$, $k$) is the index of the $j$-th neighbor of the node with index $i$ in the time slot $k$.

The 3D array *peerIds* obeys the following convention. Assume that for a given $i$ and $k$, the node $i$ has $g$ neighbors. Then the indexes of the $i$-th node neighbors are stored in the adjacent *peerIds* elements from the index ($i$, 0, $k$) to the index ($i$, $g$-1, $k$), and the *peerIds* element with the index ($i$, $g$, $k$) is set to -1 to indicate the end of the list of $i$-th node neighbors in the time slot $k$.

To enable communication among nodes, the following channels are introduced: *nodeChannels*[NoNodes] is a 1D array of channels whose elements are used as node mailboxes, and *timeSlotsMap*[NoNodes] is a 1D array of channels whose elements are used to buffer messages from faster nodes. Both channel arrays are dimensioned such that each element may hold up to (NoNodes-1)*NoTimeSlots messages (which is an upper bound on the number of messages that a node may receive from all other nodes in all time slots).

To enable returning a message (as a value) from the process IsKeyInMap, the following two items are included: *retVal*[NoNodes] is a 1D array of return values and *retChannel*[NoNodes] is a 1D array of channels whose elements may hold up to 1 message; the elements of both items correspond to individual nodes.

### *B. The Process GetMeas*

The process GetMeas, see Model 1, models the generic algorithm for universal TDM communication i.e., the PTB-FLA function getMeas. GetMeas has to parameters: *nodeId* is the node ID, and *odata* is data this node wants to send to its peers in the current time slot. If *odata* is -1 (here -1 encodes Python's None), then GetMeas increments the current time slot and terminates (line 3). Otherwise, GetMeas: uses the process SendMsgs to send its data to its peers (line 5), uses the process RecvMsgs to receive data from its peers (line 6), increments the current time slot and terminates (line 7).

Model 1. The process GetMeas.

```
01: GetMeas(nodeId, odata) =
02:   if(odata == -1) {
03:     {timeSlot[nodeId] = timeSlot[nodeId] + 1} -> Skip
04:   };
05:   SendMsgs(nodeId, odata);
06:   RecvMsgs(nodeId, odata);
07:   {timeSlot[nodeId] = timeSlot[nodeId] + 1} -> Skip;
08: SendMsgs(nodeId, odata) =
09:   SendMsgsT(0, nodeId, odata);
10: SendMsgsT(peerIdIx, nodeId, odata) =
11:   if(peerIds[nodeId][peerIdIx][timeSlot[nodeId]] != -1) {
12:     nodeChannels[peerIds[nodeId][peerIdIx][timeSlot[nodeId]]]!
          timeSlot[nodeId].nodeId.odata ->
13:     SendMsgsT(peerIdIx+1, nodeId, odata)
14: };
15: RecvMsgs(nodeId, odata) =
16:   RecvMsgsT(0, nodeId, odata);
17: RecvMsgsT(peerIdIx, nodeId, odata) =
18:   if(peerIds[nodeId][peerIdIx][timeSlot[nodeId]] != -1) {
19:     IsKeyInMap(nodeId, timeSlot[nodeId],
          peerIds[nodeId][peerIdIx][timeSlot[nodeId]]);
20:     if(retVal[nodeId] == True) {
21:       Skip
22:     } else {
23:       RecvNewMsgs(peerIds[nodeId][peerIdIx][timeSlot[nodeId]],
            nodeId, odata)
24:     };
25:     // # Unpack msg and add peerOdata to peerOdatas
```

```
26:    retChannel[nodeId]?mTimeSlot.mNodeId.mOdata ->
27:    {odataArr[nodeId][peerIdIx] = mOdata} ->
28:    RecvMsgsT(peerIdIx+1, nodeId, odata)
29:  };
30: RecvNewMsgs(peerId, nodeId, odata) =
31:   nodeChannels[nodeId]?mTimeSlot.mNodeId.mOdata ->
32:   if( (timeSlot[nodeId]!=mTimeSlot) || (peerId!=mNodeId) ) {
33:    timeSlotsMap[nodeId]!mTimeSlot.mNodeId.mOdata ->
34:    RecvNewMsgs(peerId, nodeId, odata)
35:  } else {
36:    // RecvNewMsgs shares retChannel[nodeId] with IsKeyInMap
37:    retChannel[nodeId]!mTimeSlot.mNodeId.mOdata ->
38:    Skip
39:  };
40: IsKeyInMap(nodeId, kTimeSlot, kNodeId) =
41:   {retVal[nodeId] = False} ->
42:   IsKeyInMapT(call(ccount, timeSlotsMap[nodeId]),
        nodeId, kTimeSlot, kNodeId);
43: IsKeyInMapT(noMsgs, nodeId, kTimeSlot, kNodeId) =
44:   if(noMsgs > 0) {
45:    timeSlotsMap[nodeId]?mTimeSlot.mNodeId.mOdata ->
46:    if( (kTimeSlot==mTimeSlot) && (kNodeId==mNodeId) ) {
47:      retChannel[nodeId]!mTimeSlot.mNodeId.mOdata ->
48:      {retVal[nodeId] = True} -> Skip
49:    } else {
50:    timeSlotsMap[nodeId]!mTimeSlot.mNodeId.mOdata ->
51:    Skip
52:    };
53:   IsKeyInMapT(noMsgs - 1, nodeId, kTimeSlot, kNodeId)
54:  };
```

The process SendMsgs uses the tail-recursive process SendMsgsT (lines 8-9), which in each of its passes sends the message to a single peer. SendMsgsT first checks whether it reached the end of the list of peers (line 11), and if not, SendMsgsT sends the message to the next peer (line 12), and calls itself (line 13).

The process RecvMsgs uses the tail-recursive process RcvMsgsT (lines 15-16), which in each of its passes receives a message from a single peer. RecvMsgsT first checks whether it reached the end of the list of peers (line 18), and if not, it calls the process IsKeyInMap to check whether a message from the desired peer was already received and stored in the corresponding element of the 1D array *timeSlotsMap*. If IsKeyInMap finds the message, it sets *retVal*[*nodeId*] to True and returns the message in *retChannel*[*nodeId*], and RecvMsgsT in turn skips (line 21) to line 24.

If IsKeyInMap does not find the message, RecvMsgsT calls the process RecvNewMsgs (line 23) to continue receiving incoming messages until the message from the desired peer arrives, and when it does, RecvNewMsgs returns it in *retChannel*[*nodeId*]. Further on, RcvMsgsT takes the message (line 26), extracts data sent by the desired peer and stores it into the list of data received from peers (line 27).

The process RecvNewMsgs receives a new message (line 31) and checks whether it is for the *nodeIs*'s current time slot and from the desired peer (line 32). If not, RecvNewMsgs stores the message in the *timeSlotsMap*[*nodeId*] (line 33) and calls itself (line 34) to continue receiving messages. Once RecvNewMsgs receives the desired message, it stores it in the *retChannel*[*nodeId*] (line 37) and terminates (line 38).

The process IsKeyInMap sets *retVal*[*nodeId*] to False (line 41) and calls IsKeyInMapT (line 42), which checks whether the message for the node *nodeId* and the time slot *kTimeSlot* was already sent from the node *kNodeId*. Note that *noMsgs* is the number of messages in the *timeSlotsMap*[*nodeId*], which is returned by the intrinsic function ccount.

The process IsKeyInMapT first checks whether all the messages stored in the *timeSlotsMap*[*nodeId*] were scanned (line 44). If not, IsKeyInMapT takes the next message (line 45) and checks whether it is the desired message (line 46), and if yes, IsKeyInMap stores it in the *retChannel*[*nodeId*] (line 47) and sets *retVal*[*nodeId*] to True (line 48). Otherwise, IsKeyInMap returns the message to the *timeSlotsMap*[*nodeId*] (line 50) and calls itself (line 53) to continue scanning messages stored in the *timeSlotsMap*[*nodeId*].

### *C. The System*

The process App, see Model 2, models the system i.e., a PTB-FLA application in a time frame of NoTimeSlots time slots. At the beginning, App behaves as an asynchronous interleaving of NoNodes application instances (line 2), and when the interleaving is completed, App sets *terminated* to True and terminates (line 3).

Model 2. The system.

```
01: App() =
02:   |||nodeId:{0..NoNodes-1}@AppInstance(nodeId);
03:   {terminated = True} -> Skip;
04: AppInstance(nodeId) =
05:   AppInstanceT(0, nodeId, idataArr[nodeId]);
06: AppInstanceT(appts, nodeId, odata) =
07:   if(appts < NoTimeSlots) {
08:     GetMeas(nodeId, odata);
09:     AppInstanceT(appts+1, nodeId, odata)
10:   };
```

The process AppInstance (line 4) modes an application instance i.e., a node, and it uses the tail-recursive process AppInstanceT (line 6), which captures the behaviour of an application instance during a single time stamp *appts*. If *appts* is less than NoTimeSlots (line 7), AppInstanceT first calls GetMeas (line 8) and then it calls itself for the next time stamp *appts*+1 (line 9).

## IV. FORMAL VERIFICATION IN PAT

The requested correctness properties (see Model 3) are formulated as follows: the deadlock freeness (safety property) is formulated by the single assertion in line 1, whereas the termination (liveness property) is formulated by the condition Terminated (line 2) and two assertions in lines 3 and 4. The assertion in line 3 is a weaker property, which claims that App reaches Terminated, whereas the assertion in line 4 is a stronger property, which claims that App always eventually reaches Terminated.

Model 3. The correctness properties

```
01: #assert App() deadlockfree;
02: #define Terminated (terminated == True);
03: #assert App() reaches Terminated;
04: #assert App() |= []<> Terminated;
```

The model checker PAT automatically proved the correctness properties, throughout the complete time frame (consisting of NoTimeSlots), on the completely connected communication graph (i.e., the clique), which is defined by the appropriate setup of the 3D array *peerIds*. The clique was selected because it may be considered as a worst-case graph in respect to the number of messages that are exchanged. In our future work, we plan to conduct formal verification of the correctness properties on other communication graphs, too.

## V. Related Work

In the paper [9] that is related and orthogonal to this paper, authors use celestial mechanics [4] to model the spacecraft movement, and timed automata (TA) and accompanying tool UPPAAL 4.0 to formalize and verify the first generic PTB-FLA algorithm for centralized FL. In our future work, we plan to use the same approach as a follow up of this paper.

The following three papers are more broadly related and orthogonal to this paper. The paper [10] introduces concurrency considerations at the early space mission design phases and uses UPPAAL for the mission feasibility check.

The paper [11] studies scalability of various model-checking algorithms using an arbitrarily scalable operational design describing the mode management of a satellite. While [10] and [11] consider concurrency and model checking for a single spacecraft, this paper considers communication among more spacecrafts.

The paper [12] considers a TDMA-centric inter-satellite communication specification and its verification using PVS (Prototype Verification System). The main goal of this paper was to check the well-formedness (i.e., no contradiction) of the specification and thus verify the requirements.

## VI. Conclusion

In this paper, we formally verified the correctness of the third PTB-FLA generic algorithm for the universal TDM communication, using the CSP process algebra and the model checker PAT. To the best of our knowledge, this is the first paper that formally verifies a generic algorithm of this type.

The main contributions of this paper are: (i) the CSP model of the generic algorithm for universal TDM communication, and (ii) the formulations of the requested correctness properties, namely deadlock freeness (safety property) and termination (liveness property).

The formal verification was conducted on the completely connected communication graph i.e., the clique. This is at the same time: (i) the main advantage of the paper, because the clique may be considered as a worst-case graph in respect to the number of messages that are exchanged, and (ii) the main limitation of the paper, because formal verification was conducted only on the clique.

In the future, we plan to: (i) conduct formal verification of the correctness properties on other communication graphs, too, and (ii) use celestial mechanics to model the satellite movement, and TA and UPPAAL 4.0 to formalize and verify the generic algorithm for the universal TDM communication.

## Acknowledgment

Funded by the European Union (TaRDIS, 101093006). Views and opinions expressed are however those of the author(s) only and do not necessarily reflect those of the European Union. Neither the European Union nor the granting authority can be held responsible for them.

## References

[1] M. Popovic, M. Popovic, I. Kastelan, M. Djukic, and S. Ghilezan, "A Simple Python Testbed for Federated Learning Algorithms," in Proc. of the 2023 Zooming Innovation in Consumer Technologies Conference, pp. 148-153, 2023. DOI: 10.1109/ZINC58345.2023.10173859.

[2] GitHub repo "ptbfla" [Online]. Available: https://github.com/miroslav-popovic/ptbfla. (Accessed 1 Sep 2024).

[3] TaRDIS: Trustworthy And Resilient Decentralised Intelligence For Edge Systems [Online]. Available: https://www.project-tardis.eu/ (accessed on 2 Sep 2024)

[4] M. Milankovic, Celestial Mechanics (textbook in Serbian), University of Belgrade, 1935 [Online]. Available: https://pdfcoffee.com/milutin-milankovi-nebeska-mehanikapdf-pdf-free.html (accessed on 1 Sept. 2025)

[5] S. Huang, C. Colombo, F. Bernelli-Zazzera, "Multi-criteria design of continuous global coverage Walker and Street-of-Coverage constellations through property assessment," Acta Astronautica, vol. 188, pp. 151-170, 2021. DOI: 10.1016/j.actaastro.2021.07.002.

[6] F. Caldas, C. Soares, "Machine learning in orbit estimation: A survey," Acta Astronautica, vol. 220, pp. 97–107, 2024. DOI: 10.1016/j.actaastro.2024.03.072.

[7] M. Popovic, M. Popovic, I. Kastelan, M. Djukic, and S. Ghilezan, "A Simple Python Testbed for Federated Learning Algorithms," in Proc. of the 2023 Zooming Innovation in Consumer Technologies Conference, pp. 148-153, 2023. DOI: 10.1109/ZINC58345.2023.10173859.

[8] I. Prokić, S. Ghilezan, S. Kašterović, M. Popovic, M. Popovic, I. Kaštelan, "Correct orchestration of Federated Learning generic algorithms: formalisation and verification in CSP," in: J. Kofron, T. Margaria, C. Seceleanu (eds.) Engineering of Computer-Based Systems, LNCS, Springer, Cham, vol. 14390, pp. 274-288, 2024. DOI: 10.1007/978-3-031-49252-5_25.

[9] M. Popovic, M. Popovic, M. Djukic, I. Basicevic, "Towards Formal Verification of Federated Learning Orchestration Protocols on Satellites," in Proc. 32nd IEEE Telecommunications Forum, pp. 1-4, 2024. DOI: 10.1109/TELFOR63250.2024.10819039.

[10] J. Akhundov, J. Werner, V. Schaus, A. Gerndt, "Using timed automata to check space mission feasibility in the early design phases," in Proc. of the 2016 IEEE Aerospace Conference, pp. 1-9, 2016. DOI: 10.1109/AERO.2016.7500572.

[11] P. Chrszon, P. Maurer, G. Saleip, S. Muller, P.M. Fischer, A. Gerndt, M. Felderer, "Applicability of Model Checking for Verifying Spacecraft Operational Designs," in Proc. of the 2023 ACM/IEEE 26th International Conference on Model Driven Engineering Languages and Systems (MODELS), pp. 206-216, 2023. DOI: 10.1109/MODELS58315.2023.00011.

[12] S. Gebreyohannes, R. Radhakrishnan, W. Edmonson, A. Esterline, F. Afghah, "Formalizing Inter-Satellite Communication Specification in Small Satellite System," in the Proc. of the Small Satellites, System & Services Symposium (4S), poster no. 20, pp. 1-15, 2016.